\documentclass[]{IEEEtran}
\usepackage{cite}
\usepackage[]{graphicx}
\graphicspath{{pdf/}{jpeg/}{./png/}}
\DeclareGraphicsExtensions{.pdf, .jpeg, .png}
\usepackage{amsmath}
\usepackage{array}
\ifCLASSOPTIONcompsoc
  \usepackage[caption=false, font=normalsize, labelfont=sf, textfont=sf]{subfig}
\else
  \usepackage[caption=false, font=footnotesize]{subfig}
\fi
\usepackage{url}
\usepackage{rotating}
\usepackage[OT1]{fontenc}
\usepackage{shellesc}
\usepackage[cmyk]{xcolor}
\usepackage[binary-units=true]{siunitx}
\usepackage[]{hyperref}

\usepackage{color, hyperref}
\newcommand*{\email}[1]{\href{mailto:#1}{\nolinkurl{#1}}}
\usepackage{helvet}
\usepackage[eulergreek]{sansmath}
\usepackage{dsfont}
\usepackage{amsfonts}
\usepackage[binary-units=true]{siunitx}

\begin{document}
\setcounter{footnote}{1}

\title{Machine Learning Peeling and Loss Modelling of Time-Domain Reflectometry}

\author{
    \IEEEauthorblockN{J.R.~Rinehart\thanks{\email{johnrichardrinehart@gmail.com}}%
                      \IEEEauthorrefmark{1}\IEEEauthorrefmark{2}, %
                      J.H.~B\'{e}janin
                      \IEEEauthorrefmark{1}\IEEEauthorrefmark{2}, %
                      T.C.~Fraser\IEEEauthorrefmark{2}\IEEEauthorrefmark{3}, %
                      and M.~Mariantoni%
                      \thanks{Correspondence to:\email{matteo.mariantoni@uwaterloo.ca}}%
                      \IEEEauthorrefmark{1}\IEEEauthorrefmark{2}}\\%
    \IEEEauthorblockA{\IEEEauthorrefmark{1}Institute for Quantum Computing,
    University of Waterloo, 200 University Avenue West, Waterloo, Ontario N2L 3G1, Canada}\\%
    \IEEEauthorblockA{\IEEEauthorrefmark{2}Department of Physics and Astronomy,
    University of Waterloo, 200 University Avenue West, Waterloo, Ontario N2L 3G1, Canada}\\%
    \IEEEauthorblockA{\IEEEauthorrefmark{3}Present address: Perimeter Institute
    for Theoretical Physics, 31 Caroline Street North, Waterloo, Ontario N2L 2Y5, Canada}%
}

\markboth{Machine Learning Peeling and Loss Modeling of TDR}%
{Rinehart~\MakeLowercase{\textit{et al.}}: Machine Learning Peeling and Loss
Modeling of Time-Domain Reflectometry}

\maketitle

\begin{abstract}
    A fundamental pursuit of microwave metrology is the determination of the
    characteristic impedance profile of microwave systems. Among other methods,
    this can be practically achieved by means of time-domain reflectometry~(TDR)
    that measures the reflections from a device due to an applied stimulus.
    Conventional TDR allows for the measurement of systems comprising a single
    impedance. However, real systems typically feature impedance variations that
    obscure the determination of all impedances subsequent to the first one.
    This problem has been studied previously and is generally known as
    scattering inversion or, in the context of microwave metrology, time-domain
    ``peeling''. In this article, we demonstrate the implementation of a
    space-time efficient peeling algorithm that corrects for the effect of prior
    impedance mismatch in a nonuniform lossless transmission line, regardless of
    the nature of the stimulus. We generalize TDR measurement analysis by
    introducing two tools: A stochastic \textit{machine learning} clustering
    tool and an arbitrary \textit{lossy transmission line} modeling tool. The
    former mitigates many of the imperfections typically plaguing TDR
    measurements (except for dispersion) and allows for an efficient processing
    of large datasets; the latter allows for a complete transmission line
    characterization including both conductor and dielectric
    loss.~\footnote{This work has been submitted to the IEEE for possible
    publication. Copyright may be transferred without notice, after which this
    version may no longer be accessible.}
\end{abstract}


\IEEEpeerreviewmaketitle
\section{Introduction}
  \label{sec:introduction}

\IEEEPARstart{T}{ime-domain} reflectometry~(TDR) is a microwave measurement
technique based on the response of a device under test~(DUT) to a step stimulus
incident upon it~\cite{Strickland:1970, Bryant:1993, Agilent:2013}. This
response is used to determine the characteristic impedance of the DUT along its
electrical length. Theoretically, the incident stimulus is assumed to have zero
rise time. In reality, however, this is impossible due to the finite bandwidth
of all practical systems. As a consequence, the response exhibits a nonzero rise
time that, if not accounted for, limits the measurement resolution. Another
typical analysis assumption is that the DUT contains only a single reflection
plane, whereas, in reality, many reflection planes exist. Independent of the
nature of the stimulus, multiple reflection planes induce multiple
re-reflections due to the stimulus interacting with previously encountered
reflection planes. In addition, another often neglected phenomenon of real
structures is dissipation. This results in the deterioration of the response due
to signal loss. The degree to which information regarding a lossy DUT can be
recovered is of active interest in the field of microwave metrology.

A variety of solutions have been proposed to ameliorate some of the TDR
imperfections in the lossless case. For example, in the study
of~\cite{Jong:1992}, the process of inverse scattering, known as ``peeling'',
has been introduced to microwave time-domain measurements; similarly, the works
of~\cite{Izydorczyk:2005} and \cite{Izydorczyk:2003} have proposed a peeling
algorithm in the frequency domain. An extension of the time-domain peeling
algorithm has been attempted in the study of~\cite{Liu:2013}, where a partial
solution to the problem of loss is provided. Finally, TDR utilizing arbitrary
stimuli has been investigated in the work of~\cite{Pan:2002}.

In this article, we study the conditions of applicability of the lossless
time-domain peeling algorithm in~\cite{Jong:1992} and revisit the algorithm in
the context of a \textit{machine learning technique} known as \textit{K-means
clustering} or, for brevity, K-means~\cite{Wu:2012}. We show that K-means can be
used to post-process a TDR response through data clustering, thus simplifying
the data structure and expediting peeling. The use of K-means in TDR has special
utility when considering detailed measurements that comprise a large number of
data points. Such a level of detail requires more memory to store the data and
more time to post-process it. K-means effectively equates comparable measured
samples at the benefit of reduced memory consumption and peeling computation
time. Additionally, K-means allows us to study systems with finite rise time
response as well as to efficiently account for fluctuations due to imperfections
in the stimulus, DUT, or measurement device.

Furthermore, we address lossy systems by extending the time-domain fitting
technique in~\cite{Liu:2013}, which accounts for series resistance only. We
introduce a generalized \textit{frequency-domain fitting technique}, which,
instead, accounts for \textit{both series resistance and shunt conductance}.
Provided an appropriate model for the relevant circuit parameters of interest,
our fitting technique makes it possible to determine both resistive and reactive
circuit components as arbitrary functions of frequency.

The article is organized as follows: In
Section~\ref{sec:commonly:encountered:issues}, we outline in detail many of the
common issues plaguing conventional TDR measurements and data post-processing;
in Section~\ref{sec:results}, we propose a set of solutions and assess their
efficacy by analyzing an actual measurement; finally, in
Section~\ref{sec:conclusion}, we discuss a possible combination of the tools
introduced here and outline future developments. Additionally, the article
features an appendix, Appendix~\ref{sec:a:primer:on:peeling:and:loss:fitting},
where we review the basic theory of TDR peeling and loss fitting.

A complete library including the codes used to implement our K-means TDR peeling
algorithm and circuit parameter fitting is available as an open-source software
package in~\cite{Rinehart:2017}; the codes are written in the Julia
language~\cite{Bezanson:2012:a}. The basic principles of the peeling part of the
K-means TDR peeling algorithm are
in~Appendix~\ref{sec:a:primer:on:peeling:and:loss:fitting}.

\section{Commonly-Encountered Issues}
  \label{sec:commonly:encountered:issues}

A metrologist must consider the operations that are applied by the measurement
device to the acquired data. In fact, the typical quantity reported by TDR
instruments when measuring a DUT is an \textit{improper} reflection
coefficient-like quantity~\cite{LeCroy:2007}
\begin{equation}
    \widetilde{\rho}_{0, i} = \dfrac{V^{-}[i]}{V_0} \, ,
  \label{eq:tilde:rho:0:i}
\end{equation}
where~``$0, i$'' indicates referring the impedance back to the source impedance
(with~$i \in \mathbb{N_{> 0}}$), $V^{-}[i]$ is the~$i$th sample of the reflected
voltage~\cite{Collin:2001}, and $V_0$ is the amplitude of the step stimulus. For
a DUT containing only a single reflection plane encountered at section~$\ell$,
(\ref{eq:tilde:rho:0:i}) corresponds to the reflection coefficient for~$i \ge
\ell$ (denoted as~$\rho_{i-1, i}$). However, in the presence of multiple
reflection planes, (\ref{eq:tilde:rho:0:i}) has no physical meaning and the
reflection coefficients of the DUT can only be obtained by using a peeling
algorithm.

When considering a DUT characterized by small variations of its reflection
coefficients, a peeling algorithm may not offer a significant improvement
over~(\ref{eq:tilde:rho:0:i}). In fact, in the limit of a DUT with a constant
reflection coefficient such an algorithm is not even required. Therefore, it is
reasonable to question the degree of variation requiring the use of a peeling
algorithm in order to appreciably improve measurement results. An answer to this
question can be found by analyzing the reflected voltage as a function of
repeated reflections.

\begin{figure}[t!]
    \centering
\includegraphics{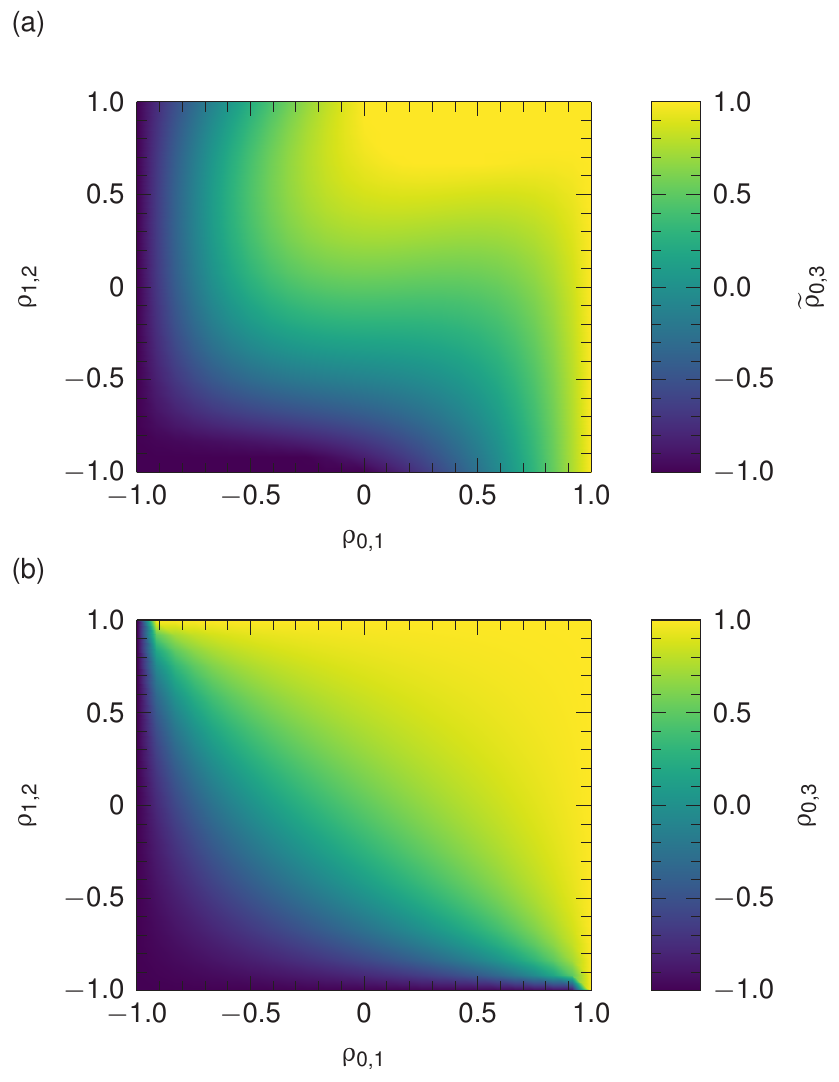}
    \caption{(Color online) Effect of prior reflection planes on the
    determination of~$\widetilde{\rho}_{0, 3}$ and $\rho_{0, 3}$~(colorbar). The
    reflection coefficient between the source and the~$1$st section of the DUT
    is~$\rho_{0, 1}$, whereas that between the~$1$st and $2$nd sections of the
    DUT is~$\rho_{1, 2}$. (a)-(b)~Plots of~(\ref{eq:tilde:rho:0:3}) and
    (\ref{eq:rho:0:i}) for~$i = 3$ and $\rho_{2, 3} = 0.5$. The strong
    discrepancy between the two plots away from the origin stresses the
    necessity of peeling when considering systems with multiple reflection
    planes.}
  \label{fig:rineh2}
\end{figure}

Consider, for example, the reflected voltage measured at time~$3 \Delta t$.
Assuming a lossless devices, the reflected voltage can be determined as
\begin{IEEEeqnarray}{rCl}
    V^{-}[3] & = & V^{+}[1] \, T_{1, 0} \, T_{2, 1} \, \rho_{2, 3} \, T_{1, 2} \, T_{0, 1} \nonumber \\
    && \negmedspace {} - V^{+}[1] \, T_{1, 0} \, \rho_{1, 2} \, \rho_{0, 1} \, \rho_{1, 2} \, T_{0, 1} \nonumber \\
    && \negmedspace {} + V^{+}[2] \, T_{0, 1} \, \rho_{1, 2} \, T_{1, 0} \nonumber \\
    && \negmedspace {} + V^{+}[3] \, \rho_{0, 1} \, ,
  \label{eq:V:-:3:simplest}
\end{IEEEeqnarray}
where~$V^{+}[i]$ is the~$i$th sample of the incident voltage; each~$\rho_{i-1,
i}$ ($T_{i-1, i}$) is the reflection (transmission) coefficient between the
$(i-1)$th and $i$th medium, with~$\rho_{i-1, i} = 1 - T_{i-1, i}$ (lossless
assumption). The na\"{i}ve approach taken to calculate~$\rho_{0, 3}$ from the
measured data is to use~(\ref{eq:tilde:rho:0:i}), resulting in the improper
coefficient
\begin{equation}
    \widetilde{\rho}_{0, 3} = \frac{V^{-}[3]}{V_{0}} \, .
\end{equation}
This result is only reasonable if all transmission coefficients are very close
to one. Equation~(\ref{eq:V:-:3:simplest}) can be rewritten in terms of only
reflection coefficients (since the device is lossless) as
\begin{IEEEeqnarray}{rCl}
    V^{-}[3] & = & V^{+}[1] \left( 1 - \rho_{0, 1}^2 \right) \left( 1 - \rho_{1, 2}^2 \right) \rho_{2, 3} \nonumber \\
    && \negmedspace {} - V^{+}[1] \left( 1 - \rho_{0, 1}^2 \right) \left( 1 - \rho_{1, 2}^2 \right) \rho_{0, 1} \nonumber \\
    && \negmedspace {} + V^{+}[2] \left( 1 - \rho_{0, 1}^2 \right) \rho_{1, 2} \nonumber \\
    && \negmedspace {} + V^{+}[3] \rho_{0, 1} \, .
  \label{eq:V:-:3}
\end{IEEEeqnarray}
By further assuming that~$V^{+}[3] = V^{+}[2] = V^{+}[1] = V_{0}$ (the case for
a step stimulus), we obtain
\begin{IEEEeqnarray}{rCl}
    \widetilde{\rho}_{0, 3} = \frac{V^{-}[3]}{V_{0}} & = & \left( 1 - \rho_{0, 1}^2 \right) \left( 1 - \rho_{1, 2}^2 \right) \rho_{2, 3} \nonumber \\
    && \negmedspace {} - \left( 1 - \rho_{0, 1}^2 \right) \left( 1 - \rho_{1, 2}^2 \right) \rho_{0, 1} \nonumber \\
    && \negmedspace {} + \left( 1 - \rho_{0, 1}^{2} \right) \rho_{1, 2} \nonumber \\
    && \negmedspace {} + \rho_{0, 1} \, .
  \label{eq:tilde:rho:0:3}
\end{IEEEeqnarray}
A general expression for the improper~$\widetilde{\rho}_{0, i}$ does not exist
due to the lack of a closed form solution for all paths (see
Appendix~\ref{sec:a:primer:on:peeling:and:loss:fitting} for a definition of
path) that contribute to the reflected voltage at discrete time~$i$. However,
the general expression for the proper~$\rho_{0, i}$, i.e., the reflection
coefficient relating the source characteristic impedance to the~$i$th
characteristic impedance section of the DUT, can be determined according
to~(\ref{eq:Z:i}) in Appendix~\ref{sec:a:primer:on:peeling:and:loss:fitting} as
\begin{equation}
    \rho_{0, i} = \frac{P^{+}_i - P^{-}_i}{P^{+}_i + P^{-}_i} \, ,
  \label{eq:rho:0:i}
\end{equation}
where
\begin{equation}
    P^{+}_i = \prod_{k=1}^i ( 1 + \rho_{k-1, k} )
\end{equation}
and
\begin{equation}
    P^{-}_i = \prod_{k=1}^i ( 1 - \rho_{k-1, k} ) \, ,
\end{equation}
with~$k \in \mathbb{N_{> 0}}$. A comparison between~(\ref{eq:tilde:rho:0:3}) and
(\ref{eq:rho:0:i}) for~$i = 3$ is shown in Fig.~\ref{fig:rineh2}~(a) and (b).

Other common issues are due to nonidealities present in the stimulus, DUT, or
measurement device that obscure information regarding the impedance profile of
the DUT. Such nonidealities include unstable stimuli, environmental sensitivity
and fabrication imprecisions of the DUT, or noisy measurement devices. For
example, the exact nature of a stimulus can affect the efficacy of peeling
algorithms significantly. In fact, the nonzero rise time of all real stimuli, if
not accounted for, may result in measurements that can be misread as having had
resulted from a lossy DUT. Additionally, the application of peeling to a
response from such a stimulus can artificially underreport the reflection
coefficients of the DUT.

\begin{figure*}[t!]
    \centering
\includegraphics{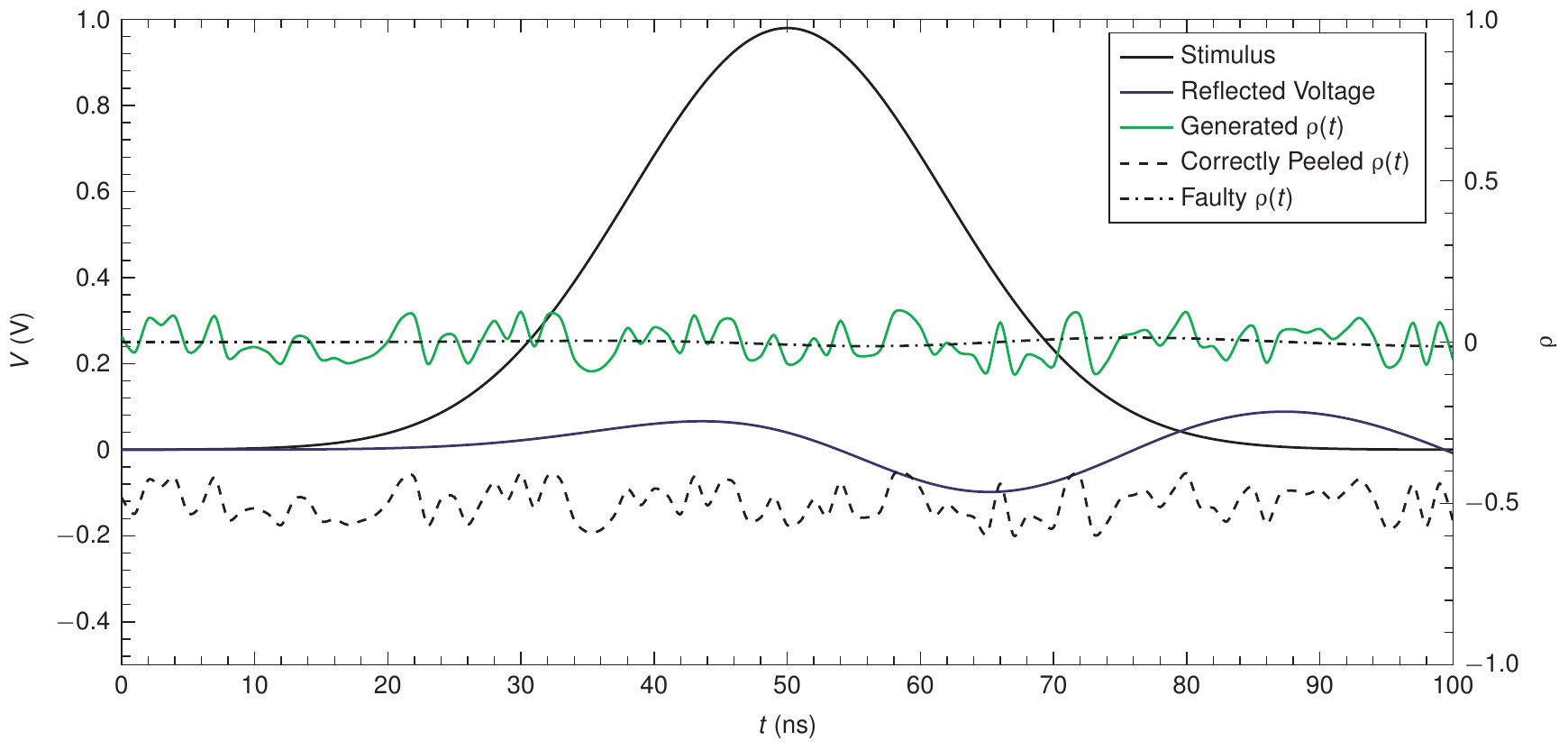}
    \caption{(Color online) Demonstration of the effect of the stimulus on the
    reflected voltage. Voltage~$V$ as a function of time~$t$~(left axis) and
    reflection coefficient~$\rho$ as a function of~$t$~(right axis). The solid
    black line is the Gaussian stimulus; the solid blue line is the reflected
    voltage obtained from the known reflection coefficients of the DUT, where
    these are shown in the solid light green (light gray) line; the dashed black
    line represents the properly peeled reflection coefficients that are
    obtained using the Gaussian stimulus and reflected voltage; the
    dashed-dotted black line represents the reflection coefficients that would
    be obtained if the stimulus were improperly assumed to be that of an ideal
    step. Note that the peeled reflection coefficients are intentionally offset
    from the known reflection coefficients for the sake of clarity; in fact,
    both these coefficients exactly coincide to within the computational
    precision of the peeling algorithm. The stark difference between the
    properly and improperly peeled reflection coefficients emphasizes the need
    for detailed knowledge of the stimulus when peeling.}
  \label{fig:rineh3}
\end{figure*}

Fig.~\ref{fig:rineh3} shows a simulated TDR measurement of a DUT with known
reflection coefficients using a Gaussian stimulus. If the peeling algorithm is
applied assuming an ideal step stimulus instead of the Gaussian stimulus, the
resulting calculated reflection coefficients are drastically different than the
known ones. This simulation is performed using the inverse of this work's
peeling algorithm implementation.

Another consideration that must be made when post-processing the reflected
voltage are the small voltage variations that appear in measurements of all
constant impedance transmission lines. These variations can be the end result of
many phenomena: Source noise; dispersion; manufacturing or setup imperfections
(e.g., trace-width variation, cable deformation, etc.); measurement noise. It is
often desired by the metrologist to ignore electrically brief small variations
in the reflected voltage in order to intentionally neglect artifacts irrelevant
to the study of the DUT. We address some of the issues outlined here in
Subsection~\ref{subsec:k:means:peeling}.

Finally, one last issue is the characterization of devices comprising
homogeneous segments of transmission line whose circuit parameters are
arbitrarily complicated functions of frequency. One simple case is that of a DUT
characterized by dielectric loss. In fact, almost all practical microwave
interconnects (microstrip, coplanar waveguide, etc.) exhibit conductance between
the signal and ground lines, resulting in dielectric loss. The study of this
loss is performed, for example, when characterizing a new substrate. As another
example, the development of superconducting microwave devices often requires
consideration of electric properties such as kinetic inductance, which is a
nonlinear function of frequency~\cite{Gross:2005}. In this case, a more general
solution than the one described in Subsection~\ref{app:subsec:loss:fitting} is
required (see Subsection~\ref{subsec:extended:loss:fitting}).

\section{Results}
  \label{sec:results}

In this section, we introduce a K-means TDR peeling technique (see
Subsection~\ref{subsec:k:means:peeling}) and an extended loss fitting technique
(see Subsection~\ref{subsec:extended:loss:fitting}).

\subsection{K-Means Peeling}
  \label{subsec:k:means:peeling}

The rapidly increasing sampling rates of microwave instruments, such as TDR
systems, will result in ever-increasing numbers of measured samples. The
state-of-the-art analog-to-digital converters achieve sampling rates of
approximately~\SI{100}{\giga S \per\second} using either conventional
electronics~\cite{Huang:2014} or superconducting
electronics~\cite{Inamdar:2009}. These sampling rates are at least ten times
larger than those used at present in standard microwave electronics and would
correspond to more than a ten-fold increase in the spatial resolution of a DUT
when employed, e.g., in TDR systems. Moreover, using frequency-domain
measurements to perform TDR indirectly, e.g., with a vector network analyzer, it
is possible to obtain an almost arbitrarily large number of samples. The
measurement of a DUT using these systems will thus require compression for
feasible post-processing of the data. Notably, compression provides further
advantage when applying state-of-the-art peeling algorithms with
runtime~$\mathcal{O}(s^2)$.

An immediate application of compression is found when considering small
time-domain fluctuations present in TDR measurements. In fact, it can be
difficult to determine whether these fluctuations are due to the stimulus,
changes in characteristic impedance, or noise in the measurement device. For a
reflected voltage containing many such fluctuations it can be desirable to
approximate them with a particular value, resulting in a compression of the
measured data. This is a remarkable example where removing unnecessary data
points using a heuristic (e.g., machine learning) approach leads to more useful
data, which are not ``contaminated'' by unwanted or unknown phenomena.

Another potential application of TDR data compression is to neglect nonzero rise
time present in the reflected voltage. Performing this operation in the context
of a nearly lossless DUT or a non-step stimulus serves to reject nonidealities
without losing important measurement information.

Compressing TDR measurements in such a way as to address the aforementioned
problems can be accomplished efficiently using a machine learning tool called
K-means~\cite{Wu:2012}. In the field of machine learning, it is often needed to
associate a set of~$s$ vectors ($s \in \mathbb{N_{> 0}}$) with a set of~$k$
vectors about which the~$s$ vectors are clustered. Such clustering algorithms
are applied heavily in the subfield of unsupervised machine learning.

In this work, we apply the K-means peeling algorithm to the TDR measurement of a
DUT in order to improve data post-processing. First, the reflected voltage data
is fed to a K-means implementation that clusters the measured voltages into
sections of any one of~$k$ values. Determining a value for~$k$ is a heuristic
process that requires some attention from the metrologist. We recommend
selecting a value of~$k$ determining segment lengths that are the smallest the
metrologist would consider. Then, the peeling algorithm is applied to the
clustered voltage samples, realizing a \textit{K-means TDR peeling} algorithm.
This algorithm is best applied to any DUT that comprises distinct homogeneous
segments of potentially unequal length. The application of K-means to the data
disregards any nonzero rise time associated with the stimulus, or response, or
both and also any small variations that may exist in the measured reflected
voltage due to noise anywhere in the measurement system.

\begin{figure}[b!]
    \centering
\includegraphics[scale=0.90]{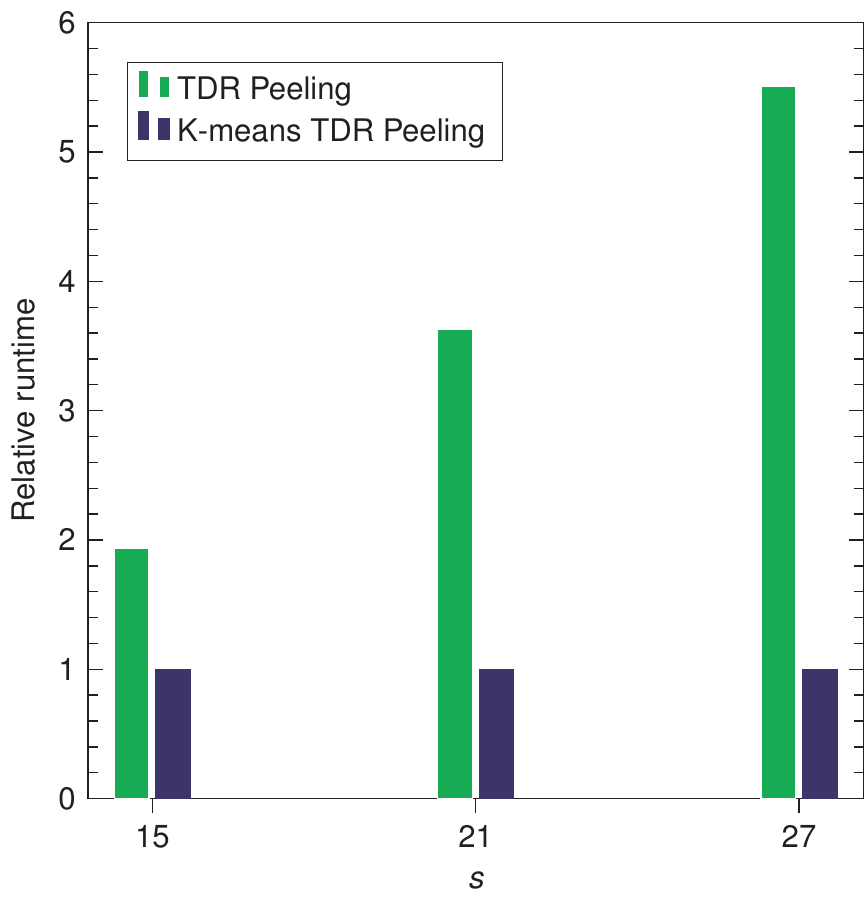}
    \caption{(Color online) Demonstration of the computational advantage of the
    K-means TDR peeling algorithm as compared to only TDR peeling~(bar chart).
    Each pair of bars compares the runtime of only TDR peeling~[light green
    (light gray) bars] to that of K-means TDR peeling~[dark blue (dark gray)
    bars] for a given set of conditions. The three pair of bars compare the
    runtimes in the context of a simulated DUT with~$s = 15$, $s = 21$, and $s =
    27$; in all three cases the DUT comprises~$s/3$ equal-length homogeneous
    sections. Note that the runtime of only the TDR peeling algorithm is scaled
    to that of K-means TDR peeling.}
  \label{fig:rineh5O}
\end{figure}

Fig.~\ref{fig:rineh5O} demonstrates the speed advantage gained by using the
K-means TDR peeling algorithm compared to simple TDR peeling. Compared in the
figure is the runtime of K-means TDR peeling given three simulated measurements,
with~$15$, $21$, and $27$~points, respectively. The runtime of K-means TDR
peeling is significantly lower than that of only TDR peeling. This is due to the
downsampling of the response. The speed-up only depends on the difference
between the number of sections before and after clustering. Although K-means
clustering algorithms typically have a time complexity of~$\mathcal{O}(s^2)$, it
has been recently shown that this threshold can be lowered
to~$\mathcal{O}(s)$~\cite{Pakhira:2014}. Therefore, in K-means TDR peeling the
complexity of the peeling part of the algorithm dominates that of K-means
clustering.

\begin{figure*}[t!]
    \centering
\includegraphics{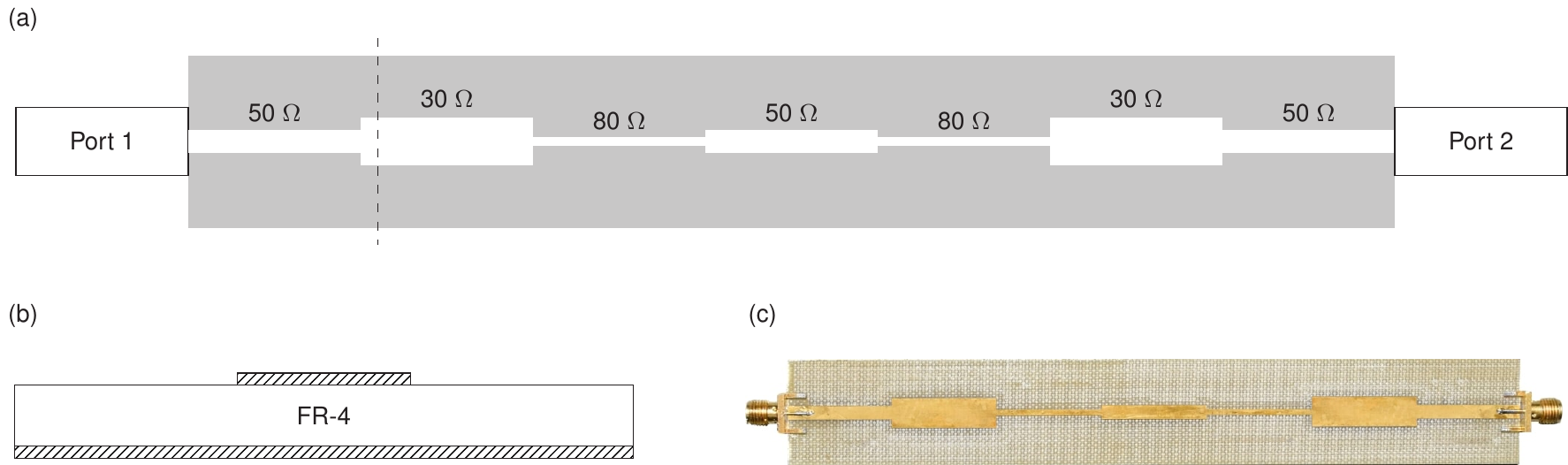}
    \caption{(Color online) Microstrip PCB layout of the DUT whose TDR
    measurement is used to test the K-means TDR peeling algorithm. The PCB is
    manufactured on a laminate comprising a~\SI{0.06}{inch}~($\approx
    \SI{1.52}{\milli\meter}$) thick FR-4 substrate and \SI{1.37}{mil}~($\approx
    \SI{34.8}{\micro\meter}$) thick copper layers on both the top and bottom
    surface of the substrate. (a)~Top view: Characteristic impedance sections of
    the DUT, $Z_1 = \SI{50}{\ohm}, Z_2 = \SI{30}{\ohm}, Z_3 = \SI{80}{\ohm}, Z_4
    = \SI{50}{\ohm}, Z_5 = \SI{80}{\ohm}, Z_6 = \SI{30}{\ohm}$, and $Z_7 =
    \SI{50}{\ohm}$, with PCB layout drawn to scale (substrate in gray and
    conductor in white). Each section is~\SI{1}{inch}~(\SI{2.54}{\centi\meter})
    long and, for reference, the~\SI{50}{\ohm} sections are~$\approx
    \SI{131}{mil}$ ($\approx \SI{3.33}{\milli\meter}$) wide. Port~$1$ is used as
    the port for the TDR measurement; port~$2$ is terminated in
    a~$\SI{50}{\ohm}$~load. Practical limitations prohibit manufacturing
    arbitrarily large or small impedance sections. Small impedances are limited
    by the available size of the PCB and practical considerations, while large
    impedances are limited by the minimum trace width attainable given a
    particular fabrication process. Using this laminate and microstrip geometry,
    a trace width greater than~\SI{1}{inch} is required in order to obtain
    a~\SI{10}{\ohm}~impedance. The narrowest impedance is bounded by
    the~$\SI{4/1000}{inch}$~(\SI{101.6}{\micro\meter}) minimum trace width
    imposed by our PCB milling machine. This corresponds to~$\approx
    \SI{168}{\ohm}$ for our laminate; however, working close to the milling
    machine limits results in significant errors. For this reason, a wider trace
    width is chosen, which determines a lower impedance than the maximum
    allowable. The dashed black line indicates the plane of the cross section
    shown in (b). (b)~Side view: Cross section of the DUT at~$Z_2$ (heights not
    to scale, widths to scale), i.e., corresponding to the dashed black line
    in~(a). The conductor layers are hatched; the substrate is shown in white.
    (c)~Photograph of the microstrip PCB as fabricated; ports~$1$ and $2$ are
    connected to a subminiature type A, or SMA, PCB female connector.}
  \label{fig:rineh4}
\end{figure*}

To gauge the efficacy of the K-means TDR peeling algorithm in a practical
context, a microstrip stepped impedance DUT is designed and fabricated on a
printed circuit board~(PCB). A scaled layout of the DUT (as well as its material
properties) is supplied in Fig.~\ref{fig:rineh4}. The design of this PCB is
based on two criteria:
\begin{enumerate}
    \item Include many consecutive large changes in impedance;
    \item ensure the round-trip delay of the smallest segment is much greater
    than the sampling rate of the measurement device.
\end{enumerate}

The first criterion, in particular, is important to ensure that multiple
re-reflections significantly impact the measured voltage. In fact, small
impedance variations would result in re-reflections that are too small to be
detected with the TDR system used in our experiments, which is from Teledyne
LeCroy, model WaveExpert~100H.
    \footnote{The TDR oscilloscope features an electrical sampling module
    with~\SI{20}{\giga\hertz} bandwidth and a TDR step generator, model~ST-20.
    The generated signal is a voltage square wave characterized by a nominal
    pulse rise time of~\SI{20}{\pico\second}, amplitude
    of~\SI{250}{\milli\volt}, pulse width of~\SI{300}{\nano\second}, and pulse
    repetition rate of~\SI{1}{\mega\hertz}.}
The input and output impedance of this TDR system, the impedance of Port~1 and
Port~2 in Fig.~\ref{fig:rineh4}, is~$Z_0 \approx \SI{50}{\ohm}$.
    \footnote{A single measurement comprising a single terminating impedance (in
    this case, the~\SI{50}{\ohm} load of Port~2) is sufficient to characterize
    the DUT. Given the nature of peeling, measurements using other load
    impedances would not affect the determination of the impedance profile of
    the DUT.}
Thus, the PCB impedance values are chosen such as to vary significantly relative
to~$Z_0$. Given the ease of manufacturing microstrip impedances
between~\SI{20}{\ohm} and \SI{100}{\ohm} on a standard copper-clad FR-4
substrate, the design shown in Fig.~\ref{fig:rineh4} is chosen. \footnote{The
microstrip PCB is fabricated using a circuit board plotter from LPKF Laser \&
Electronics AG, model LPKF ProtoMat~S$103$.}

\begin{figure*}[t!]
    \centering
\includegraphics{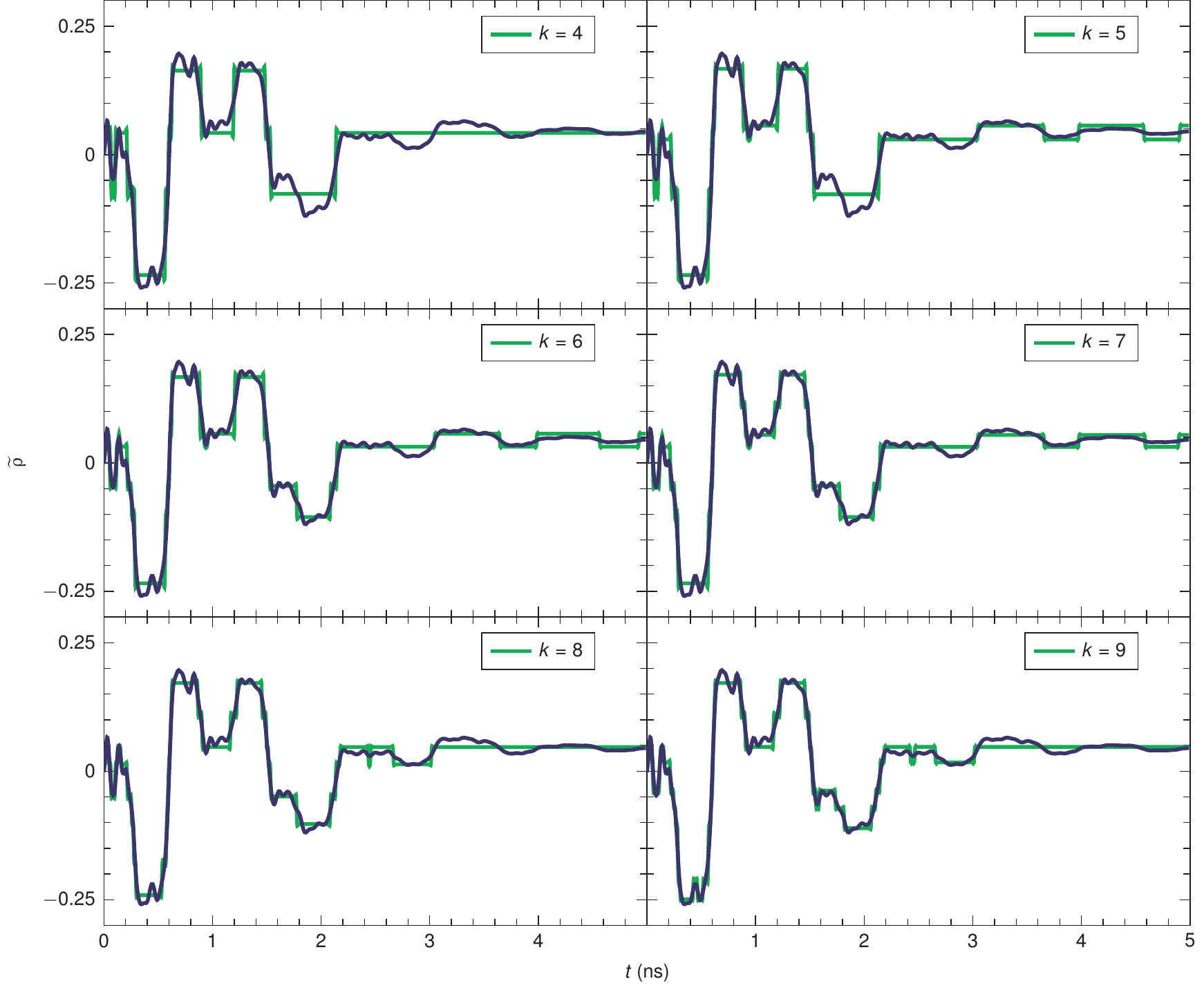}
    \caption{(Color online) K-means clustering applied to the DUT measurement
    for a variety of cluster sizes~$k = 4, 5, \ldots, 9$. The vertical axis
    reports the improper reflection coefficient~$\widetilde{\rho}$ (i.e., before
    peeling) as a function of time~$t$. Measured data is shown in dark blue
    (dark gray) and clustered data in light green (light gray). The cluster size
    is a heuristic quantity determined by the metrologist and depends on the
    form of the measured data. This demonstrates the effect of different cluster
    sizes on the resulting post-processed data, hence affecting peeling results.
    For this reason, it is recommended to cluster the data with a set of varying
    cluster sizes in order to determine the sensitivity of peeled reflection
    coefficients to the number of clusters.}
  \label{fig:rineh5}
\end{figure*}

Our PCB design has the advantage of being symmetric, so that TDR measurements
can be performed from both ports of the DUT for comparison. Additionally, the
impedance varies significantly between different impedance steps exacerbating
the effect of multiple re-reflections interfering at the source. The time-domain
measurement of the microwave PCB and the result of clustering the data for a
number of different cluster sizes are shown in Fig.~\ref{fig:rineh5}.

\begin{figure*}[t!]
    \centering
\includegraphics{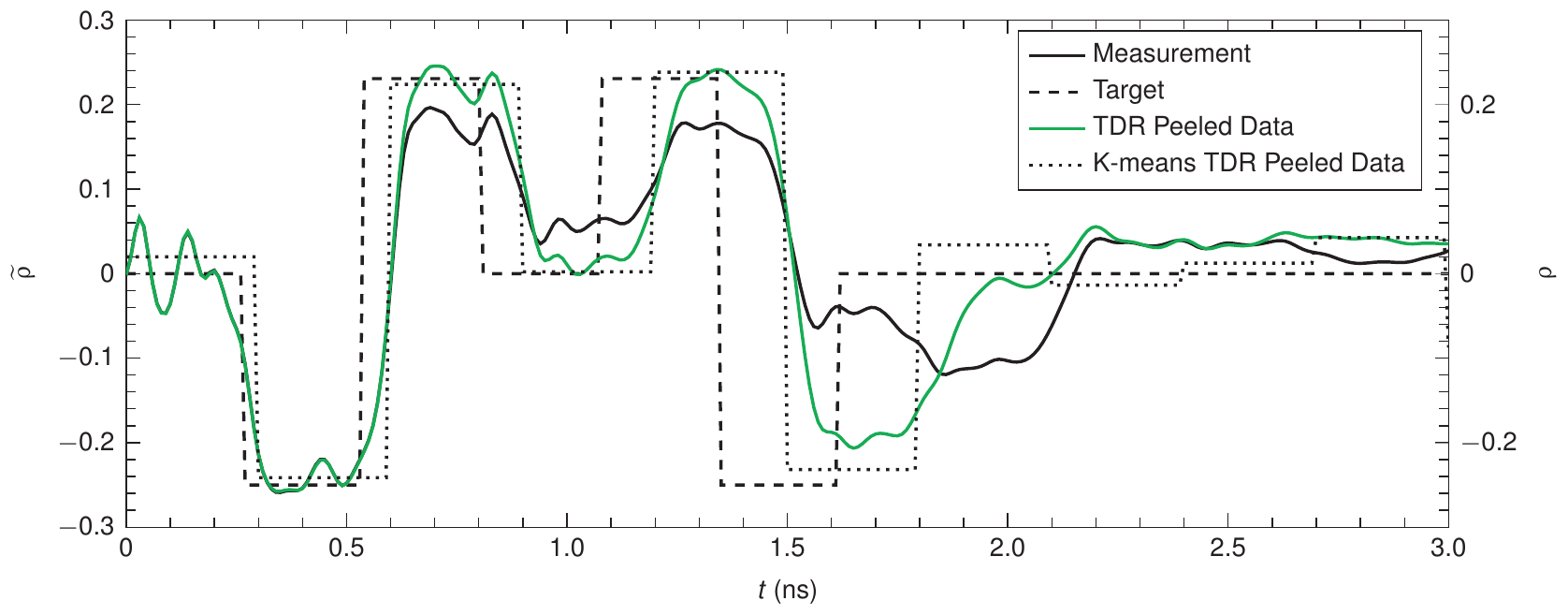}
    \caption{(Color online) Demonstration of the K-means TDR peeling algorithm
    applied to the measurement of the DUT in Fig.~\ref{fig:rineh4}. Improper
    reflection coefficient~$\widetilde{\rho}$ as a function of time~$t$ (left
    axis) and proper reflection coefficient~$\rho$ as a function of~$t$ (right
    axis). The solid black line represents the measured improper reflection
    coefficients as determined by~(\ref{eq:tilde:rho:0:i}); the dashed black
    line represents the designed reflection coefficient of each section; the
    solid light green (light gray) line is the standard TDR peeling applied to
    the measured data; the dotted black line is the K-means TDR peeling applied
    to the measured data. The effect of microstrip dispersion is visible in each
    of the three lines involving measured data (see
    Section~\ref{sec:conclusion}).}
  \label{fig:rineh6}
\end{figure*}

Using the measurement values as reported by the TDR system, we can immediately
identify the pitfalls of using the improper reflection coefficient
of~(\ref{eq:tilde:rho:0:i}). The reflection coefficient (occurring at~$\approx
\SI{0.3}{\nano\second}$) is reported to be~$\approx - 0.25$, as designed. The
second value, however, is reported to be less than~$0.17$, which is
significantly smaller than the designed reflection coefficient of~$\approx
0.23$. It is in such situations that K-means TDR peeling harbors most potential.
The efficaciousness of K-means clustering combined with peeling is shown in
Fig.~\ref{fig:rineh6}, which demonstrates that the determination of all
reflection planes is improved compared to the improper (na\"{i}ve) case. Note
that the disagreement between the theoretical result (target) and the K-means
peeled measurement can be attributed to the dispersion and loss of the DUT,
which are not accounted for.

While K-means peeling seemingly involves \textit{ad hoc} manipulation of
measured data, this can serve to the benefit of the metrologist. Measured data
unavoidably contains artifacts unrelated to the properties of the DUT, which
originate from both the source and measurement instruments. The effect of these
artifacts on the measured data is usually assumed to be negligibly small.
However, due to the nonlinear nature of the peeling algorithm, a reflected
voltage signal containing small artifacts can radically affect the peeled
measurement. Applying K-means clustering prior to peeling has the potential to
average the effect of zero-mean noise introduced by both the source and meter.
This potentially improves the data used as input to a peeling algorithm.
Comparing the results of peeling both with and without the application of
K-means clustering provides a mechanism for determining the degree of error
introduced in the measurement as a result of a nonideal source and meter.

\subsection{Extended Loss Fitting}
  \label{subsec:extended:loss:fitting}

Dissipation in electrical systems typically arises from radiative, conductor,
and dielectric loss, with radiative loss being a negligible effect in many
practical scenarios. All these effects depend on frequency, as in the case of
conductor losses due to the skin effect~\cite{Collin:2001}. The reactive
components of a DUT, i.e., inductance and capacitance, also depend on frequency,
as in the case of the kinetic inductance in superconductors~\cite{Gross:2005}.
Given appropriate theoretical frequency-domain models of each circuit parameter,
it is possible to determine their contribution to the response. We find these
parameters by extending the fitting technique described in~\cite{Liu:2013}.
While in that study the fitting method has been performed in the time domain and
accommodates only a subset of all circuit parameters (see
Section~\ref{sec:commonly:encountered:issues}), our method is performed in the
frequency domain and accommodates all circuit parameters which, additionally,
can be arbitrary functions of frequency.

The characteristic impedance of a transmission line is given
by~(\ref{eq:Z:characteristic}) in
Appendix~\ref{sec:a:primer:on:peeling:and:loss:fitting}. Thus, the knowledge
of~$r(f)$, $l(f)$, $g(f)$, and $c(f)$ makes it possible to determine the
time-domain response of the DUT. However, even if a detailed knowledge of the
geometrical and material properties of the DUT is provided, it is impractical to
construct a complete circuit-parameter model. Typical approximations to all
circuit parameters are given by~(\ref{eq:r:f}) and (\ref{eq:l:f:c:f}) in
Appendix~\ref{sec:a:primer:on:peeling:and:loss:fitting}, as well as
\begin{equation}
    g(f) = 2 \pi f \, c(f) \, \epsilon_{\textrm{r}}^{\prime\prime} \, ,
\end{equation}
where~$c$ is the per-unit-length capacitance and
$\epsilon_{\textrm{r}}^{\prime\prime}$ the imaginary part of the relative
electric permittivity of the DUT substrate. These assumptions result in the five
parameter frequency-domain model of the characteristic impedance
\begin{align}
    Z ( f ; r_{\textrm{dc}} , r_{\textrm{s}} , l_0 , & c_0 , \epsilon_{\textrm{r}}^{\prime\prime} ) = \nonumber \\
    & \sqrt{\frac{r_{\textrm{dc}} + [ 1 + j \, \textrm{sgn}(f) ] \, r_{\textrm{s}} \sqrt{\left| f \right|} + j \, 2 \pi f \, l_0} {2 \pi f \, c_0 \, \epsilon_{\textrm{r}}^{\prime\prime} + j \, 2 \pi f \, c_0}}
  \label{eq:Z:five:parameters:normal}
\end{align}
that applies to a wide variety of lossy transmission lines.

\begin{figure*}[t!]
    \centering
\includegraphics[scale=0.95]{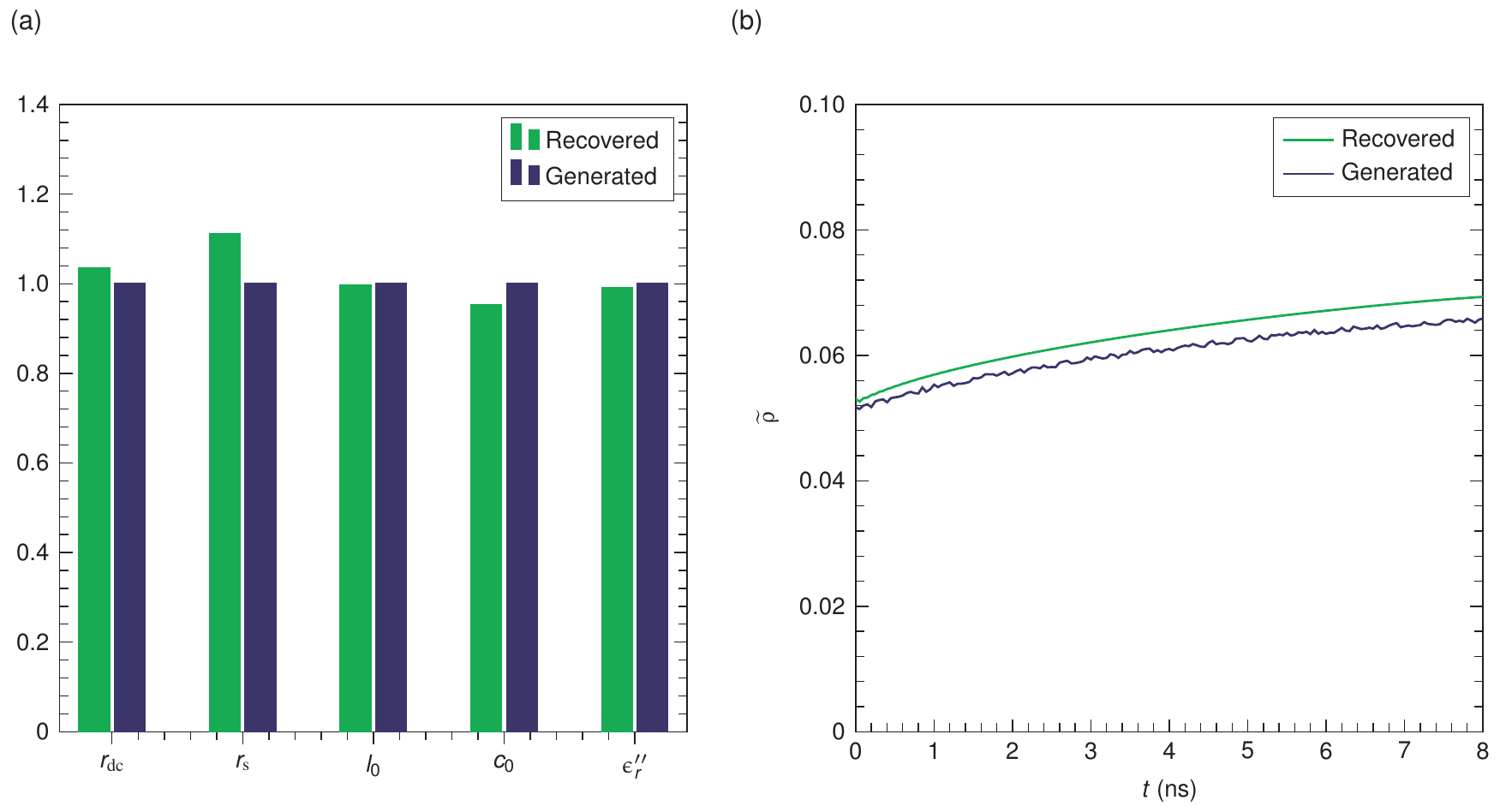}
    \caption{(Color online) Demonstration of the loss fitting technique
    described in Subsection~\ref{subsec:extended:loss:fitting} using the five
    parameter model of~(\ref{eq:Z:five:parameters:normal}). A DUT of
    parameters~$r_{\textrm{dc}} = \SI{290}{\milli\ohm\per\meter}$,
    $r_{\textrm{s}} = \SI{45}{\micro\ohm\per\hertz\tothe{1/2}\per\meter}$, $l_0
    = \SI{300}{\nano\henry\per\meter}$, $c_0 = \SI{100}{\pico\farad\per\meter}$,
    and $\epsilon_{\textrm{r}}^{\prime\prime} = 1.05 \cdot 10^{-2}$ is used in
    the simulation, with~$s = 162$.  This simulation is performed using an
    inverse process to that described in
    Subsection~\ref{subsec:extended:loss:fitting}. Gaussian noise is added to
    the time-domain data in order to simulate the effect of source and
    measurement noise. (a)~Comparison of the generated (i.e., simulated)
    parameters [dark blue (dark gray)] to the recovered parameters~[light green
    (light gray); bar chart]. The recovered parameters are scaled to the
    generated parameters. (b)~The dark blue (dark gray) line is the time-domain
    response of the generated DUT; the light green (light gray) line is the
    simulation of the time-domain response of a device synthesized with the
    recovered parameters.}
  \label{fig:rineh7}
\end{figure*}

Determining the parameters in~(\ref{eq:Z:five:parameters:normal}) requires
fitting them to a sequence of measured data points in the frequency domain. In
order to assess the efficacy of this fitting scheme we:
\begin{enumerate}
    \item Simulate the time-domain response of a DUT to a step stimulus using
    the inverse of our peeling algorithm;
    \item introduce noise associated with both the stimulus and measurement
    device using a pseudorandom number generator;
    \item perform a discrete Fourier transform of the time-domain response to
    determine the frequency-domain response of the DUT;
    \item fit the frequency-domain response obtained in~3) using a nonlinear
    least-squares fitting routine based on the Levenberg-Marquardt algorithm.
\end{enumerate}
This procedure serves to recover all five parameters
in~(\ref{eq:Z:five:parameters:normal}). The time-domain response of the
recovered DUT and the recovered parameters are shown in Fig.~\ref{fig:rineh7}.

As another example of the broad applicability of this fitting scheme, we
consider the case of a superconductor at a temperature~$T$ below its critical
temperature~$T_{\textrm{c}}$, $T < T_{\textrm{c}}$. Using the surface impedance
of a superconductor in lieu of~(\ref{eq:r:f}) and (\ref{eq:l:f:c:f}), we can
write the characteristic impedance model of a homogeneous section of
superconducting transmission line segment as~\cite{Gross:2005}
\begin{align}
    Z ( f , T , & T_{\textrm{c}} ; \lambda_{\textrm{L}} , \sigma_0 , \lambda_0 , c_0 , \epsilon_{\textrm{r}}^{\prime\prime} ) = \nonumber \\
    & \sqrt{\dfrac{2 \pi^2 f^2 \mu_0^{2} \lambda_{\textrm{L}}^3 \dfrac{\sigma_{0}}{L} \left(\dfrac{T}{T_{\textrm{c}}}\right)^4 + j \, 2 \pi f \, \dfrac{\mu_0 \lambda_{\textrm{L}}}{L} + j \, 2 \pi f \, l_0}{2 \pi f \, c_0 \, \epsilon_{\textrm{r}}^{\prime\prime} + j \, 2 \pi f c_0}} \, ,
  \label{eq:Z:five:parameters:super}
\end{align}
where~$\mu_0$ is the magnetic constant, $\lambda_{\textrm{L}}$ the London
penetration depth of the superconductor, $\sigma_0$ the conductivity of the
normal state conductor, and $L$ the length of the superconducting transmission
line segment. A similar fitting scheme as the one enumerated above can be used
to find all five frequency-dependent circuit parameters
in~(\ref{eq:Z:five:parameters:super}).

\section{Conclusion}
  \label{sec:conclusion}

In this article, we introduce a TDR peeling algorithm based on the K-means
clustering method that provides a significant improvement over known
alternatives such as in~\cite{Jong:1992, Izydorczyk:2005, Izydorczyk:2003}. We
show examples where traditional peeling algorithms fail and present a practical
case study to benchmark the K-means TDR peeling and demonstrate its superiority
over traditional peeling. While we apply this machine learning method to the
case of TDR measurements, other related techniques can be used in a wider
variety of measurement post-processing tasks. This is particularly important in
the era of big data, where large datasets allow for more precise and accurate
measurements of devices at the cost of increased memory requirements and
post-processing runtime.

In addition, we present a method to perform an extended characterization of
devices using a fitting scheme that makes it possible to account for a DUT
described by a parametrized frequency-domain model. This fitting scheme can be
used in conjunction with the K-means TDR peeling algorithm in order to assess
devices comprising multiple homogeneous segments, as proposed
in~\cite{Liu:2013}. We show frequency-domain models for both normal conductors
and superconductors.

It is worth noting that, in some instances, it may be necessary to account for
dispersion when applying the K-means TDR peeling algorithm. An example of this
can be seen in Fig.~\ref{fig:rineh6}, where the algorithm is applied to a
microstrip transmission line. We will address this issue in future work since it
affects a variety of microwave systems. Additionally, a TDR peeling algorithm
involving measurements with aperiodic samples would provide large benefit when
combined with K-means clustering and should be the focus of further
investigations. Finally, a generalized peeling technique accommodating lossy
devices that does not require nonlinear fitting is desired and merits further
attention.

\section*{Acknowledgment}

This research was undertaken thanks in part to funding from the Canada First
Research Excellence Fund~(CFREF), as well as the Discovery and Research Tools
and Instruments Grant Programs of the Natural Sciences and Engineering Research
Council of Canada~(NSERC) and the Ministry of Research and Innovation~(MRI) of
Ontario. We acknowledge our fruitful discussions with Eric Bogatin, Thomas
G.~McConkey, and A.~Hamed Majedi; JRR acknowledges his fruitful discussions with
Behrooz Semnani.

\appendices

\section{A Primer on Peeling and Loss Fitting}
  \label{sec:a:primer:on:peeling:and:loss:fitting}

In this section, we review the state-of-the-art on TDR peeling algorithms (see
Subsection~\ref{app:subsec:peeling}) and loss fitting techniques (see
Subsection~\ref{app:subsec:loss:fitting}).

\subsection{Peeling}
  \label{app:subsec:peeling}

\begin{figure*}[t!]
    \centering
\includegraphics[scale=0.90]{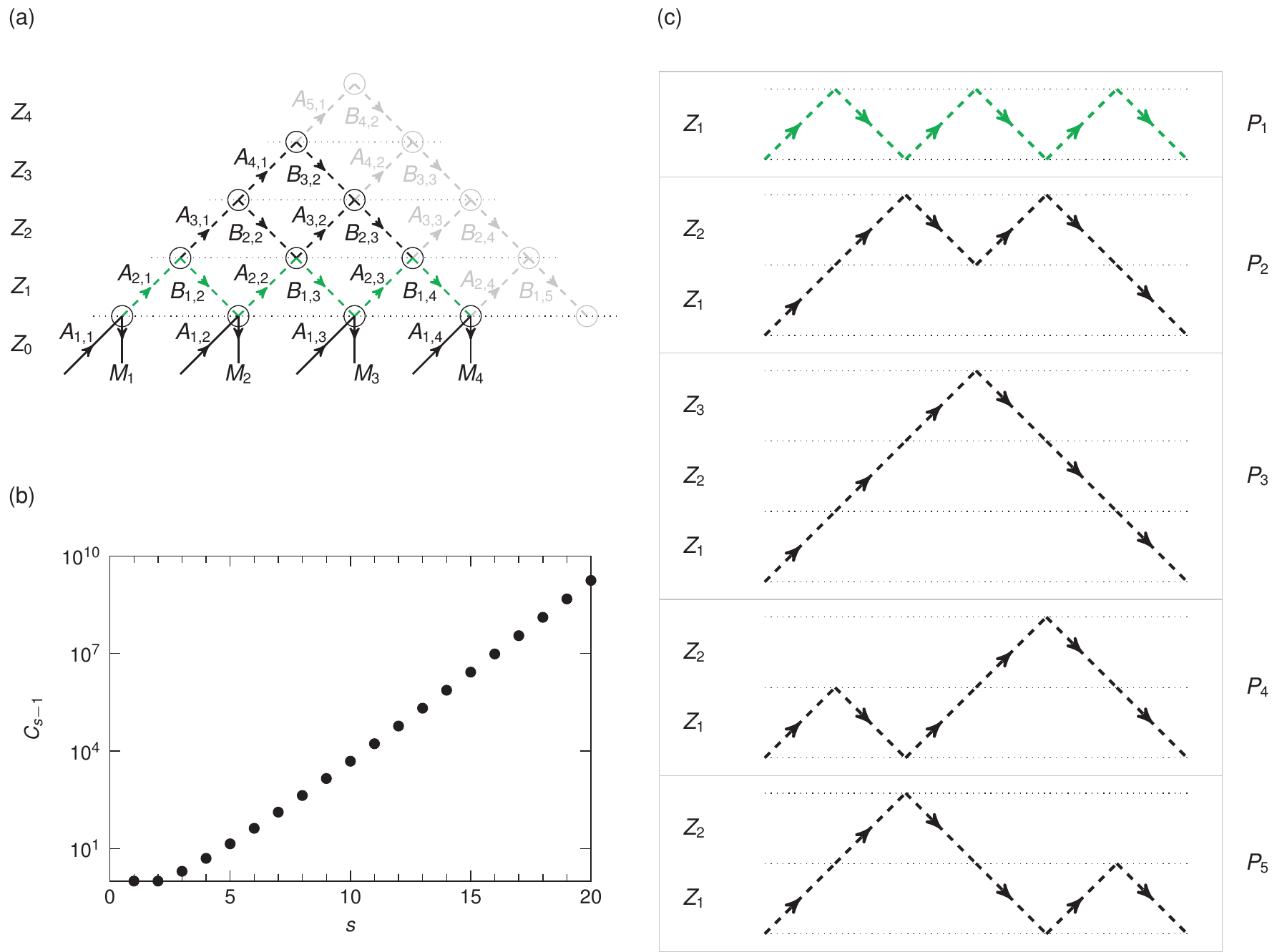}
    \caption{Diagrammatic representation of all possible paths a stimulus can
    take in a DUT. This representation is key to understanding all recursive
    peeling algorithms. (a)~The forward- and backward-oriented segments are
    labelled by~$A_{i, j}$ and $B_{i, j}$, respectively, where~$i$ indicates the
    depth into the DUT and~$j$ the location of each segment with respect to the
    upper right diagonal starting at each~$M_j$ in~(a). The voltage weight of
    each~$A_{1, j}$ is the stimulus at time step~$j$ and that of~$M_j$ is
    the~$j$th sample of the measured reflected voltage. Each~$Z_i$ is the~$i$th
    characteristic impedance section of the DUT, delimited by horizontal dotted
    lines. These lines indicate reflection planes due to the discretization of
    the sampled response [similarly in~(c)]. Solid segments are those whose
    voltage weights are directly accessible to the metrologist, whereas dashed
    segments indicate internal paths whose weights are indirectly accessible by
    means of peeling. Open circles represent nodes, where each node is
    associated with at least one transmission or reflection event, or both. The
    light gray segments and nodes depict an extension of the DUT. The light
    green (light gray) path is highlighted for comparison with the similarly
    highlighted path in~(c). (b)~Scaling of the number of paths~$C_{s-1}$ with
    the number of measured samples~$s$. (c)~All~$C_3 = 5$ possible paths that
    the stimulus can take to arrive at the source in time~$t_4 = 4 \Delta t$.
    Each path is labelled by~$P_k$, with~$k = 1, 2, \ldots, 5$.}
  \label{fig:rineh8}
\end{figure*}

A TDR measurement of a DUT with multiple reflection planes requires
post-processing in order to accurately characterize the DUT. The stimulus used
in these measurements is a travelling voltage wave, which is scattered within
the DUT due to the presence of reflection planes. Each scattered component
arrives at the measurement plane at delayed times, in accordance with its
scattering history, and there interferes with other scattered components. Thus,
there exists a one-to-one correspondence between a \textit{path} through the DUT
and a \textit{reflected voltage component}. The path \textit{weight}, which is
the reflected voltage amplitude for a path, is determined by the stimulus as
well as the value of each encountered reflection coefficient. All of these
components combine to form the total measured voltage response, which is the
signal used to assess the DUT. This process can be inverted using peeling
algorithms, which are designed to determine the reflection coefficients of the
DUT from the measured reflected voltage~\cite{Jong:1992, Izydorczyk:2005,
Izydorczyk:2003}.

Fig.~\ref{fig:rineh8}~(a) shows a schematic representation of a DUT
characterized by multiple reflection planes, where the scattered voltage can
take different paths through the DUT. Each path carries information regarding
the reflection planes that it encounters when travelling from the source to the
measurement device. These paths are characterized by forward- and
backward-oriented \textit{segments} labeled as~$A_{i, j}$ and $B_{i, j}$,
respectively, with~$i, j \in \mathbb{N_{> 0}}$.

In general, time-domain reflectometers attempt to sample the reflected voltage
at a periodic time interval~$\Delta t$, which determines an implicit
discretization of the DUT into a set of sections whose electrical properties can
be determined by the measured voltage (see~\cite{Agilent:2013} for detailed TDR
measurement information). Each measured sample must, \textit{a priori}, be
assumed to be associated with one reflection plane, even if it is of zero value.
Suppose there are~$s$ voltage samples of an~$N$-section DUT whose length is such
that~$s \ge N$ ($N, s \in \mathbb{N_{> 0}}$). In this case, it can be shown that
the number of paths associated with the~$s$ planes is the Catalan number
for~$(s-1)$~\cite{Kim:2016},
\begin{equation}
    C_{s-1} = \frac{(2s-2)!}{s!(s-1)!} \, .
  \label{eq:C:s-1}
\end{equation}
Fig.~\ref{fig:rineh8}~(b) illustrates the strong functional dependence
of~(\ref{eq:C:s-1}) on the number of planes. Note that for~$s = 150$, $C_{s-1}$
is greater than the estimated number of particles in the universe. Storing the
information regarding all of the paths is impractical considering the scaling
behavior of~(\ref{eq:C:s-1}). Notably, this is unnecessary since the~$C_{s-1}$
paths can be constructed using only information from~$s$ reflection planes. For
example, Fig.~\ref{fig:rineh8}~(c) indicates that~$5$ paths each of~$6$ segments
can be computed with only~$4$ reflection coefficients. Thus, the solution is not
to solve for all the weights of all paths, but to recursively compute the values
of the reflection coefficients by computing the values of each segment once.
This is the essence of state-of-the-art time-domain peeling algorithms.

In order to determine the reflected voltage contribution from each path it is
necessary to consider the weight of each segment used to construct it. These
weights are determined by the nature of the impedance profile. Each impedance
variation introduces a possible reflection or transmission event, or both,
forming a node in Fig.~\ref{fig:rineh8}~(a). A numerical implementation of the
recursive time-domain peeling algorithm in the lossless case is
in~\cite{Rinehart:2017}. This algorithm generates an array~$\rho[i]$
corresponding to~$\rho_{i-1, i}$ and operates in runtime~$\mathcal{O}(s^2)$,
with~$s$ the number of measured reflected voltage samples.

An efficient approach to peeling utilizes two facts:
\begin{enumerate}
    \item The~$i$th acquired sample is the first sample with information
    regarding the~$i$th reflection plane;
    \item only one path involves the~$i$th reflection plane at the~$i$th time
    step.
\end{enumerate}

Having had determined the proper reflection coefficients of the DUT by peeling,
the impedance profile of the DUT can be determined by
\begin{equation}
    Z_i = Z_{i-1} \frac{1 + \rho_{i-1, i}}{1 - \rho_{i-1, i}} \, .
  \label{eq:Z:i}
\end{equation}

\subsection{Loss Fitting}
  \label{app:subsec:loss:fitting}

The most general microwave model of a DUT involves four per-unit-length
quantities that are each a function of frequency~$f$ (in units of
hertz)~\cite{Collin:2001}: The series resistance~$r(f)$; the series
inductance~$l(f)$; the shunt capacitance~$c(f)$; the shunt conductance~$g(f)$.
Prior work has established a time-domain fitting method to determine these
quantities, with the exception, however, of~$g(f)$~\cite{Liu:2013}. This partial
fitting is performed under the assumptions:
\begin{enumerate}
    \item
        \begin{equation}
            r(f) = r_{\textrm{dc}} + r_{\textrm{s}} [ 1 + j \, \textrm{sgn} ( f ) ] \sqrt{\left| f \right|} \, ,
          \label{eq:r:f}
        \end{equation}
        where~$r_{\textrm{dc}}$ is the dc series resistance per unit length,
        $r_{\textrm{s}}$ a skin effect-related resistance per unit length and
        per root hertz, $j^2 = -1$, and ``$\textrm{sgn}$'' the signum function
        (cf.~\cite{Johnson:1993});
    \item
        \begin{equation}
            l(f) = l_0 \quad \text{and}
            \quad
            c(f) = c_0 \, ,
          \label{eq:l:f:c:f}
        \end{equation}
        where~$l_0$ and $c_0$ are a constant inductance and capacitance per unit
        length, respectively;
    \item $g(f) = 0$.
\end{enumerate}
The characteristic impedance of any section of a DUT can be expressed as
\begin{equation}
    Z \left( r(f) , l(f) , g(f) , c(f) \right) = \sqrt{\frac{r(f) + j \, 2 \pi f \, l(f)}{g(f) + j \, 2 \pi f \, c(f)}} \, .
  \label{eq:Z:characteristic}
\end{equation}
Considering a DUT satisfying the three aforementioned assumptions,
(\ref{eq:Z:characteristic}) can be simplified to
\begin{multline}
    Z ( f ; r_{\textrm{dc}} , r_{\textrm{s}} , l_0, c_0 ) = \\
    \sqrt{\frac{r_{\textrm{dc}} + r_{\textrm{s}} [ 1 + j \, \textrm{sgn}(f) ] \sqrt{\left| f \right|} + 2 \pi f \, l_0}{j \, 2 \pi f \, c_0}} \, .
\end{multline}
Assuming~$r(f)$ to be small relative to~$\sqrt{l_0/c_0}$, this expression can be
expanded in a Maclaurin series to first order yielding
\begin{multline}
    Z ( f ; r_{\textrm{dc}} , r_{\textrm{s}} , l_0 , c_0 ) \simeq \\
    \sqrt{\frac{l_0}{c_0}} \left( 1 + \frac{1}{2} \frac{r_{\textrm{dc}} + r_{\textrm{s}} [ 1 + j \, \textrm{sgn}(f) ] \sqrt{\left| f \right|}}{j \, 2 \pi f \, l_0} \right) \, .
  \label{eq:approx:Z:characteristic}
\end{multline}
Multiplying~(\ref{eq:approx:Z:characteristic}) by the Fourier transform of a
step function and then performing an inverse Fourier transform results in a
time-domain function that can be fitted using a suitable optimization routine.
This procedure makes it possible to obtain the four constant
parameters~$r_{\textrm{dc}} , r_{\textrm{s}} , l_0$, and $c_0$. In order to
implement the fitting procedure, it is necessary to measure a sequence of
time-domain reflected voltage values obtained from a DUT with constant circuit
parameters over its electrical length, i.e., a DUT that is spatially homogeneous
with respect to these parameters. When combining this procedure with the peeling
algorithm, it is possible to characterize a nonhomogeneous DUT that contains a
set of homogeneous sections.

\bibliographystyle{IEEEtran}
%\bibliography{rineh_bibliography}

\begin{IEEEbiography}
[{\includegraphics[width=1in, height=1.25in, clip, keepaspectratio]{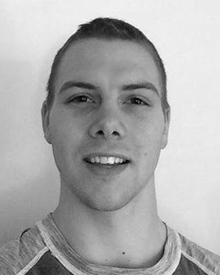}}]
{John R.~Rinehart} received the Bachelor of Science degree in electrical
engineering \textit{summa cum laude} in~$2013$ from Gonzaga University, Spokane,
USA. From~$2011$ to $2012$, he was with~LHC2 Inc., where he helped develop
antenna receiver architectures. In the summer of~$2012$, he worked alongside
researchers from the Martinis Group at the University of California, Santa
Barbara, USA, on the implementation of a packaging technique for superconducting
quantum bits~(qubits). Since~$2013$, he is enrolled in the quantum information
graduate program at the Institute for Quantum Computing~(IQC) at the University
of Waterloo~(UW), Waterloo, Canada, as a PhD candidate in physics. In Waterloo,
Rinehart is part of a team developing both classical and quantum technologies
for quantum computing with superconducting devices. John R.~Rinehart
co-authored~$3$~peer-reviewed articles and is a student member of the~IEEE.
\end{IEEEbiography}

\begin{IEEEbiography}
[{\includegraphics[width=1in, height=1.25in, clip, keepaspectratio]{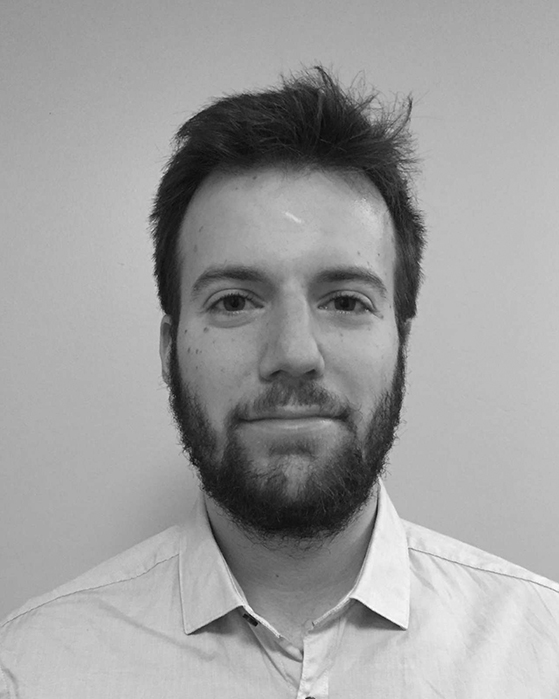}}]
{J\'{e}r\'{e}my H.~B\'{e}janin} received the Bachelor of Science degree in
physics in~$2013$ from McGill University, Montreal, Canada. Since then, he is
enrolled in the quantum information graduate program at the IQC at UW, Waterloo,
Canada, as a PhD candidate in physics. In Waterloo, he is part of a team
developing both classical and quantum technologies for quantum computing with
superconducting devices. His current research focuses are on superconducting
qubits and resonators, optimal control theory, as well as quantum many-body
physical models. B\'{e}janin co-authored~$3$~peer-reviewed articles.
\end{IEEEbiography}

\begin{IEEEbiography}
[{\includegraphics[width=1in, height=1.25in, clip, keepaspectratio]{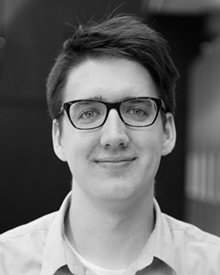}}]
{Thomas C.~Fraser} received the Bachelor of Science degree in Honours
Co-operative Mathematical Physics (Astrophysics Specialization) in~$2017$ from
the UW, Canada. In~$2016$, he was awarded the UW Mike Lazaridis Scholarship in
Theoretical Physics at the Perimeter Institute~(PI) for Theoretical Physics,
Canada. In~$2017$, he received the Governor General's Academic Medal (Silver
Medal) for academic excellence. Since~$2017$, he has been pursuing a master's
degree as a student of the Perimeter Scholars International~(PSI) program at~PI
with interest in quantum information. In particular, Fraser's current research
is concerned with systematically identifying the resources provided by entangled
quantum states through the perspective of causal inference.
\end{IEEEbiography}

\vfill
\begin{IEEEbiography}
[{\includegraphics[width=1in, height=1.25in, clip, keepaspectratio]{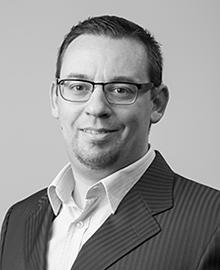}}]
{Matteo Mariantoni} (MScEng'04) received a doctorate degree in physics
(Dr.~Rer.~Nat.) \textit{summa cum laude} in~$2009$ from the Technical University
of Munich, Germany. In Munich, he performed experiments on propagating quantum
microwave fields and worked on scalable quantum computing architectures.
In~$2009$, he was awarded the Elings Prize Fellowship in Science of the
California NanoSystems Institute at the University of California, Santa Barbara,
USA. During his postdoctoral research in Santa Barbara, Mariantoni's
implementation of a quantum memory and processor on a single chip using a
``quantum von Neumann architecture'' was ranked as one of the top~$10$
breakthroughs for~$2011$ by Physics World. In December~$2012$, he moved to
the~IQC at the UW, Waterloo, Canada, as an Assistant Professor in the Department
of Physics and Astronomy, where he leads a team developing both classical and
quantum technologies for quantum computing with superconducting devices. He was
awarded an Alfred P.~Sloan Research Fellowship in~$2013$ as well as the Ontario
Early Researcher Award and the Kavli Fellowship from the National Academy of
Sciences in~$2014$. Mariantoni co-authored~$35$~peer-reviewed articles totaling
more than~$3500$ citations and authored one book.
\end{IEEEbiography}

%\newpage
%\vfill
\end{document}